\documentclass[twocolumn]{aa}
\usepackage{graphicx}
\begin{document}
   \title{Center-to-Limb Variation of Solar Line Profiles as a Test  
	 of NLTE Line Formation Calculations}


   \author{Carlos Allende Prieto
          \inst{1,2},
	Martin Asplund
          \inst{3}, 
         \and
          Pe\~na Fabiani Bendicho\inst{2,4}
          }

   \offprints{C. Allende Prieto}

   \institute{McDonald Observatory and Department of Astronomy,
              University of Texas, Austin, TX 78712-1083, USA\\
        \and
             Instituto de Astrof\'{\i}sica de Canarias, E-38200, La Laguna, 
		Spain\\        
         \and
	Research School of Astronomy and Astrophysics,
	Mt. Stromlo Observatory, Cotter Rd., Weston,
	ACT 2611, Australia\\   
         \and
	     Departamento de Astrof\'{\i}sica, Universidad de La Laguna,
		E-38206, La Laguna, Spain\\
\email{callende@astro.as.utexas.edu,martin@mso.anu.edu.au,pfb@ll.iac.es}
             }

   \date{}

   \abstract{

We present new observations of the center-to-limb variation of spectral
lines in the quiet Sun. 
Our long-slit spectra are corrected for scattered
light, which amounts to 4--8 \% of the continuum intensity, by comparison
with a Fourier transform spectrum of the disk center. 
Different spectral lines exhibit different behaviors, 
depending on their sensitivity to the physical conditions in the photosphere
and the range of depths they probe as a function of the observing angle,
providing  a rich database to test  
models of the solar photosphere  and  line formation.
We examine the
effect of inelastic collisions with neutral hydrogen in NLTE line
formation calculations of the oxygen infrared triplet, and the 
Na I $\lambda6160.8$ line.
Adopting a classical one-dimensional theoretical model atmosphere, 
we find  that the sodium transition, formed in  higher layers, is much more 
effectively thermalized by hydrogen collisions than the high-excitation
oxygen lines. This result appears as a simple consequence of the decrease of 
the ratio N$_{\rm H}/$N$_{\rm e}$ with depth in the solar photosphere. 
The center-to-limb variation of the selected lines
 is studied both under LTE and NLTE conditions.
In the NLTE analysis, inelastic collisions with hydrogen atoms are
considered with a simple approximation  
or neglected, in an attempt to
test the validity of such approximation.
For the sodium line studied,  the best agreement between 
theory and observation happens when NLTE is considered and inelastic 
collisions with hydrogen are neglected in the rate equations.
The analysis of the oxygen triplet benefits from a very detailed
calculation using an LTE three-dimensional model atmosphere and NLTE 
line formation. The $\chi^2$ statistics favors
 including hydrogen collisions with
the  approximation adopted, but the oxygen abundance derived in that
case is significantly higher than the value derived from OH infrared
transitions.

   \keywords{Sun: photosphere --
		Line: formation --
                Line: profiles
               }
   }

   \maketitle
%

\section{Introduction}

Abundance analyses of late-type stars commonly rely on the assumption
of Local Thermodynamical Equilibrium (LTE) and a plane-parallel homogeneous
structure. Even though in many cases the derived abundances may
be actually similar to the real photospheric abundances, there are well-known
instances when departures from LTE and inhomogeneities can introduce
significant systematic errors. Consequently, efforts have been directed
toward improving the modeling techniques. 
One of the most significant obstacles in the way of performing reliable 
Non-LTE (NLTE) calculations is the incompleteness of the necessary atomic data.
Because the calculations tend to be involved, and rely on approximations
to account for rates that have never been measured in a laboratory or
cannot be calculated reliably, it is often the case that 
results cannot be directly accepted without extensive testing and 
thorough comparison with observations. 

Common tests that have been put into practice
 are spectroscopic observations of
 profiles of lines with well-known damping constants, 
abundance determinations for stars in a cluster  (expecting all members
to show the same chemical composition)
or a solar abundance analysis for non-volatile elements 
(anticipating that  the photospheric
abundances and those found in chondrites are identical).
The spatially resolved solar disk offers the possibility to survey a
 range of formation depths for the continuum, and also for any
 given spectral line.
The center-to-limb variation of spectral lines
constitutes a particularly interesting test for elements like oxygen, which
is depleted in meteorites (see e.g. Sedlmayr 1974; Kiselman 1991). 
Furthermore, this type of observations may turn very useful to
distinguish between the thermalizing effect of inelastic
collisions with electrons and 
 hydrogen atoms, as the ratio N$_{\rm e}$/N$_{\rm H}$ 
increases by more than one order of magnitude between 
$\log \tau_{5000} = -2$ and $0$. 

Inspection of the available literature reveals a flagrant scarcity of
high quality observations of the center-to-limb variation of line profiles.
Early work did not benefit from digital 
detectors 
(e.g. M\"uller \& Mutschlecner 1964; M\"uller, Baschek, \& Holweger 1968).
The extended study by Balthasar (1988) includes more than one hundred lines, but
only reports equivalent widths and a few parameters related to  
 the amount and shape of the line asymmetry. 
Equivalent widths, however, neglect much of 
the available information in a line
profile. Equivalent widths ignore, for example, whether
 changes in the damping wings  or the core
of a line dominate the line strength variation, 
or even if they cancel each other to produce an equivalent width 
nearly independent of the position on the disk. 
More recent work exists, but it is limited to a small number of
spectral regions (e.g. Ambruoso et al. 1992; Brandt \& Steinegger 1998; 
Grigoryeva \& Turova 1998; Langhans \& Schmidt 2002; Stenflo et al. 1997).

We have obtained 
observations of  a number of key lines for which reliable damping 
constants are available.  In Section \ref{obs} we report our measurements and
the data reduction, in particular a procedure to subtract the scattered
light. In Section \ref{clv}  we study  the cases of
the infrared oxygen triplet and the Na I $\lambda6161$ lines in the
context of a classical model atmosphere. In Section  \ref{3d}, the case
of the oxygen triplet is reanalyzed with a time-dependent three-dimensional
model of the solar photosphere. The paper concludes with some reflections 
about the results and suggestions for future work.

\section{Observations and data reduction}
\label{obs}

Solar observations of the center-to-limb variation of several spectral lines
were carried out in October 22-23, 1997, with the Gregory Coud\'e 
Telescope (GCT) and its Czerny-Turner echelle 
spectrograph (Kneer et al. 1987; Kneer \& Wiehr 1989) at  the 
Observatorio del Teide (Tenerife, Spain). 
This telescope was moved to Tenerife and refurbished 
in 1985, after more than  20 years  of operations 
in Locarno (Switzerland). 
It was dismantled in 2002 to leave room for new instrumentation. 
The spectrograph was operated in orders 6--9 
with a slit width between 50 and 150 $\mu$m (0.4--1.3 arcsec)
achieving an  estimated FWHM ($\equiv \delta\lambda$) resolving power 
$R\equiv \lambda/\delta\lambda$ in the range 57,000--240,000. 

\begin{table}
      \caption[]{Observations.}
         \label{t1}
     $$ 
         \begin{array}{p{0.3\linewidth}ccc}
            \hline
            \noalign{\smallskip}
            Central $\lambda$    &   {R'} & 
		S ~{\mathrm{(Sct. light)}} & {\mathrm{max}}(\theta) \\
             \noalign{\smallskip} (\AA)   &  ^{\mathrm{(a)}}  & 
		 & {\mathrm{(deg)}} \\
            \noalign{\smallskip}
            \hline
            \noalign{\smallskip}
5245   &    56000  &     0.054 & 75 \\
5300   &    176000 &    0.046  & 80 \\
6122   &    77000  &   0.065   & 80 \\
6162   &    148000 &    0.066  & 75 \\
6166   &    77000  &   0.066   & 75 \\
6200   &    206000 &    0.080  & 75 \\
7585   &    176000 &    0.048  & 80 \\
7770   &    86000  &   0.058   & 80 \\
          \noalign{\smallskip}
            \hline
         \end{array}
     $$ 
\begin{list}{}{}
\item[$^{\mathrm{(a)}}$] $R$' is the resolving power measured relative to that 
of the FTS spectra; see text
\end{list}
   \end{table}

\begin{figure}
\centering
{\includegraphics[width=6.cm,angle=90]{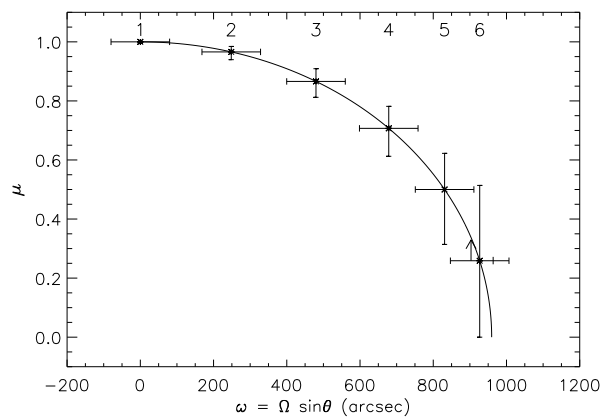}}
\caption{Correspondence between the location and extent of 
the slit for each position in $\mu \equiv 
\cos \theta$ and the angle on the sky along the observations line  
from the center of the disk 
$\omega \equiv \Omega \sin \theta$, where $\Omega$ is the
solar angular radius which is assumed to be 959.5 arcsec. The
positions of the center of the slit are marked with asterisks. For 
position \# 6, part of the slit was outside the solar disk; 
the center of the illuminated part of the slit 
is located at the tip of the arrow, and the
right limit is marked with a vertical line at $\omega = \Omega$.}
\label{slit}
\end{figure}

We secured spectra for 8 spectral setups in 6 different 
positions across the solar disk, as summarized in Table 1.  A ninth setup
was centered  at about 7610  \AA,
to obtain an estimate of the amount of  scattered light from  
saturated telluric O$_2$ lines.
The exact slit location was always chosen to avoid active regions.
Ten consecutive equal-length exposures were obtained
at each position for each setup. The exposure time varied,
depending on the setup,  between 0.5 and 2.0 seconds. 
Positions \#1 to \#5 were always at heliocentric angles $\theta$ = 
0, 15, 30, 45, and 60 degrees 
($\mu \equiv \cos \theta$ = 1.00, 0.97, 0.87, 0.71, and 0.50)
 along a straight line crossing the center of the solar disk.
Position \#6 was also selected along
the same direction, sometimes at $\theta=$75 degrees and others at 80 degrees
($\mu$ = 0.26 or 0.17).
The slit length covered approximately 205 arcsec,
 but the field  was truncated by the CCD size to $\approx$ 160 arcsec.
In the last position, part of the slit was outside the solar disk,
and therefore when the center of the slit was at $\theta=$75 and 80 deg,
the center of the illuminated slit was at 70 and 72 deg, respectively, or
approximately $\mu \simeq 0.32$. Fig. \ref{slit} shows the slit
extent both in angular size on the sky and in $\mu$ for each position,
assuming a solar angular radius of $\Omega=$ 959.5 arcsec
 (Allende Prieto et al. 2003a).
In what follows we neglect limb-darkening within the slit
length, assigning any observed intensity to the location of 
the center of the illuminated part of the slit.

\begin{figure}[t!]
\centering
{\includegraphics[width=6.cm,angle=90]{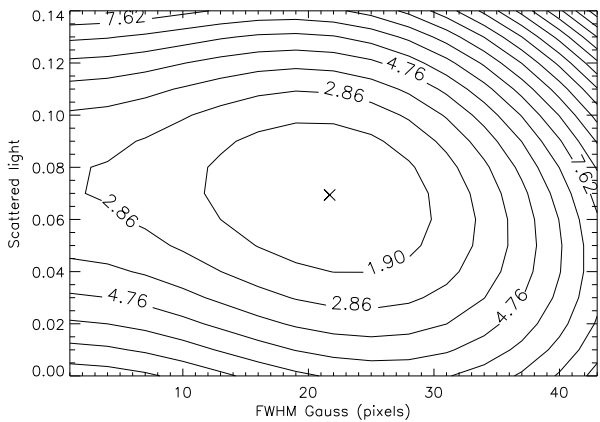}}
\caption{Contour plot showing the reduced $\chi^2$,
normalized to unity at its minimum, 
as a function of the relative resolving power 
and the amount of scattered light for
the $\mu=1$ setup centered around the Ca I $\lambda6162$ line.}
\label{chi2}
\end{figure}

No calibration lamps were available. The detector was 
a 1024$^2$ CCD with negligible dark current  (1.4 ADU/sec/pixel). 
The CCD frames were bias subtracted, and rotated. The rotation
corrects small
tilts between the  direction of the spectral dispersion and the CCD
that ranged between 0.5 and 0.7 degrees, as derived from the drift of 
the centers of several spectral lines in the spatial direction. 
Pixel-to-pixel variations and weak fringes 
  were automatically smeared out in the spatially-averaged spectra 
at each position.
Wavelength calibration of the disk-center spectra
was carried out by fitting a 1st to 3rd order
polynomial  (depending on the number of available spectral lines)
 to the wavelengths of the line centers, as measured in 
the Brault \& Neckel atlas (1987; see Neckel 1994). The wavelengths
were adopted from Allende Prieto \& Garc\'{\i}a L\'opez (1998), or
measured afresh when missing from their list. The same dispersion solution was
applied to the observations at other positions. 
Finally, a velocity shift relative to the center-of-the-disk spectra
(accounting mainly for solar rotation, Earth's motion, and the limb effect) were
determined by cross-correlation and applied to the spectra at other 
positions.
Although all the
observations for a given spectral setup were acquired consecutively, 
the time interval between the observations at the disk center and at the
limb was generally significant, and therefore 
we deem our wavelength accuracy
insufficient to study the limb effect.

The spectrum of O$_2$ lines at about 7600 \AA\ (A band) 
confirmed the suspicion that 
scattered light was substantial. The presence
of scattered light is also apparent when comparing our spectra
with the Fourier transform (FT) spectrum at the center of the disk in the
Brault \& Neckel atlas.
 Fourier transform spectrographs  (FTS's)
are not susceptible to scattered light in the same sense 
as grating  spectrographs, but non-linearities in the detector can cause systematic
errors in the zero-point of the intensity scale (see the discussion in
Kurucz et al. 1984). 
Scattered light leaves a characteristic distortion on the spectrum. On visual
inspection, for a given spectral resolution, the cores of the lines appear filled. 
This effect is noticeably different from a reduction in spectral resolution. 
As the spectrum of the quiet Sun averaged over a large
area is believed to be extremely stable, two instrumental factors  
are mainly responsible for the differences between
 the FT spectrum of Brault \& Neckel 
and our center-of-the-disk spectra: resolving power, and the amount
of scattered light. Therefore, based on the well-supported 
expectation that the amount of scattered light in the FTS atlas is much less 
that in our GCT spectra, we used the FTS data as a template to correct the 
scattered light\footnote{The O$_2$ lines of the 
A-band, which are  presumed opaque, show $\simeq$ 0.5 \% of the
pseudo-continuum intensity in the FT spectrum,
 while our spectra indicate $>2$ \%.}.

\begin{figure}[t!]
\centering
{\includegraphics[width=6.3cm,angle=90]{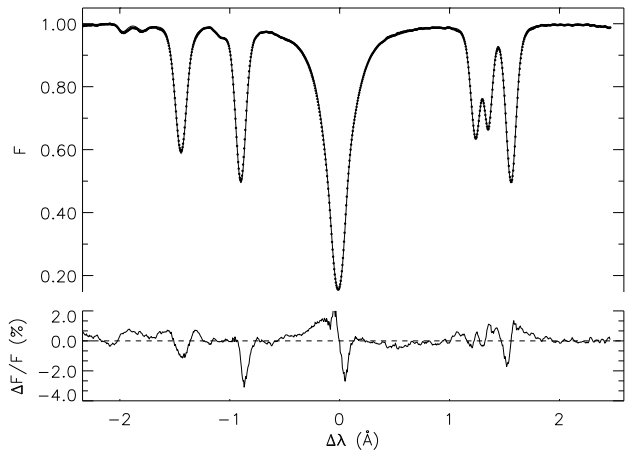}}
\caption{Comparison of the spectra at the center of the disk in the region
around Ca I $\lambda6162.2$ from
Brault \& Neckel (1987; convolved with a Gaussian; solid line) and the Gregory 
Coud\'e Telescope (after removal of scattered light; filled circles).}
\label{f2}
\end{figure}

For each of the setups, we compared our center-of-the-disk spectra
with the FTS atlas to derive the amount of scattered light and
the resolving power ($R'$ relative to the FTS 
atlas\footnote{The relative resolving power $R'$ can be approximately 
converted to absolute resolving power knowing that $R_{\rm FTS} \simeq 
4 \times 10^5$, and $R= (1/R'^2-1/R_{\rm FTS}^2)^{-1/2}$.}
) that would lead to the
best agreement between the two. Our analysis was based on 
modeling the instrumental profile of the GCT spectra as a Gaussian, and
the scattered light as a constant fraction of the continuum flux for each 
setup. 
The comparison required the GCT spectra to be first continuum corrected.
This was accomplished by using a 6th-order polynomial and a series
of clipping iterations. Then a search was performed to determine the
best-fitting values for the FWHM of the Gaussian representing the
instrumental profile, and $S$, the amount of scattered light expressed as
a fraction of the pseudo-continuum intensity.

\begin{figure}[t!]
\centering
{\includegraphics[width=6.3cm,angle=90]{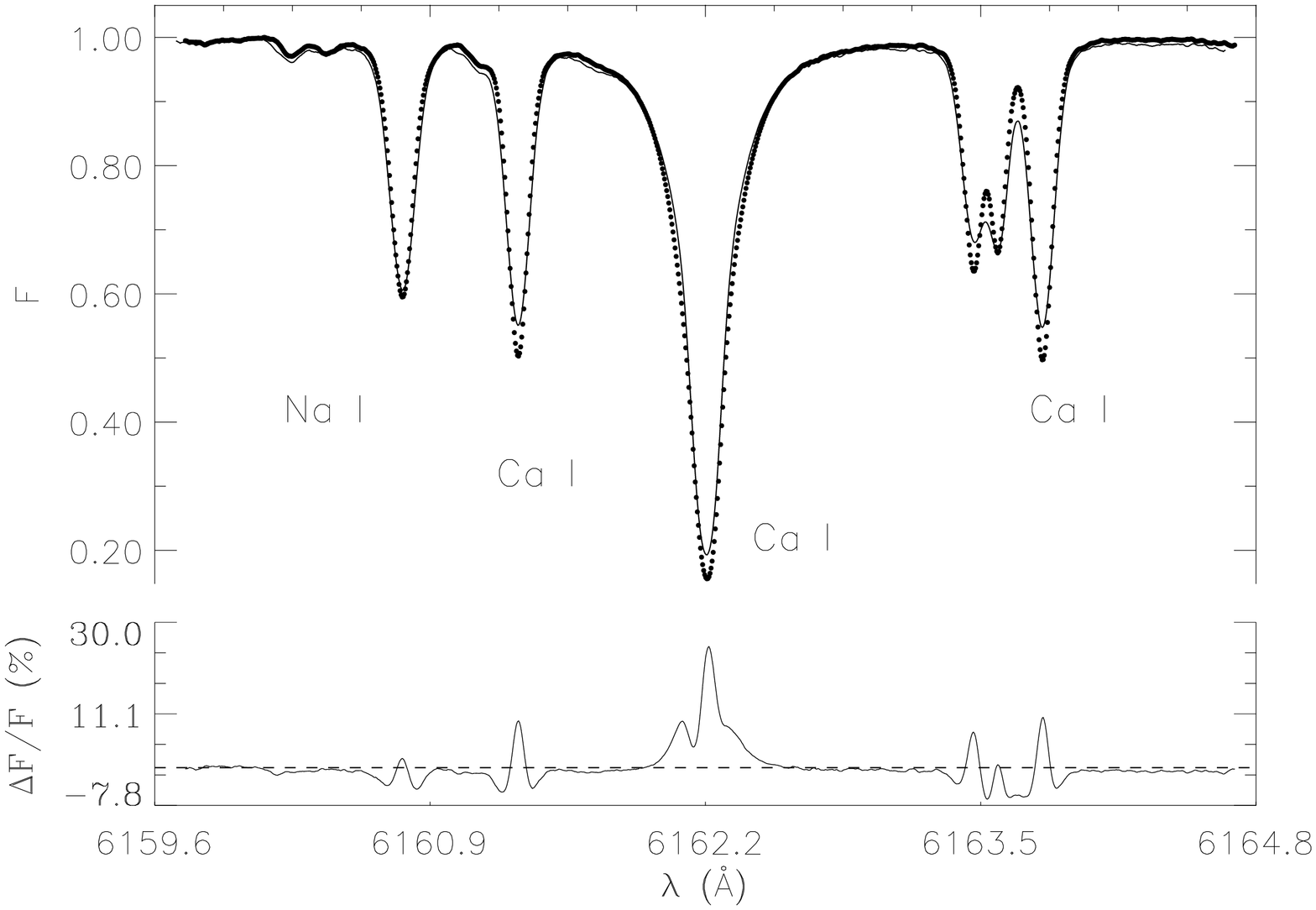}}
{\includegraphics[width=6.3cm,angle=90]{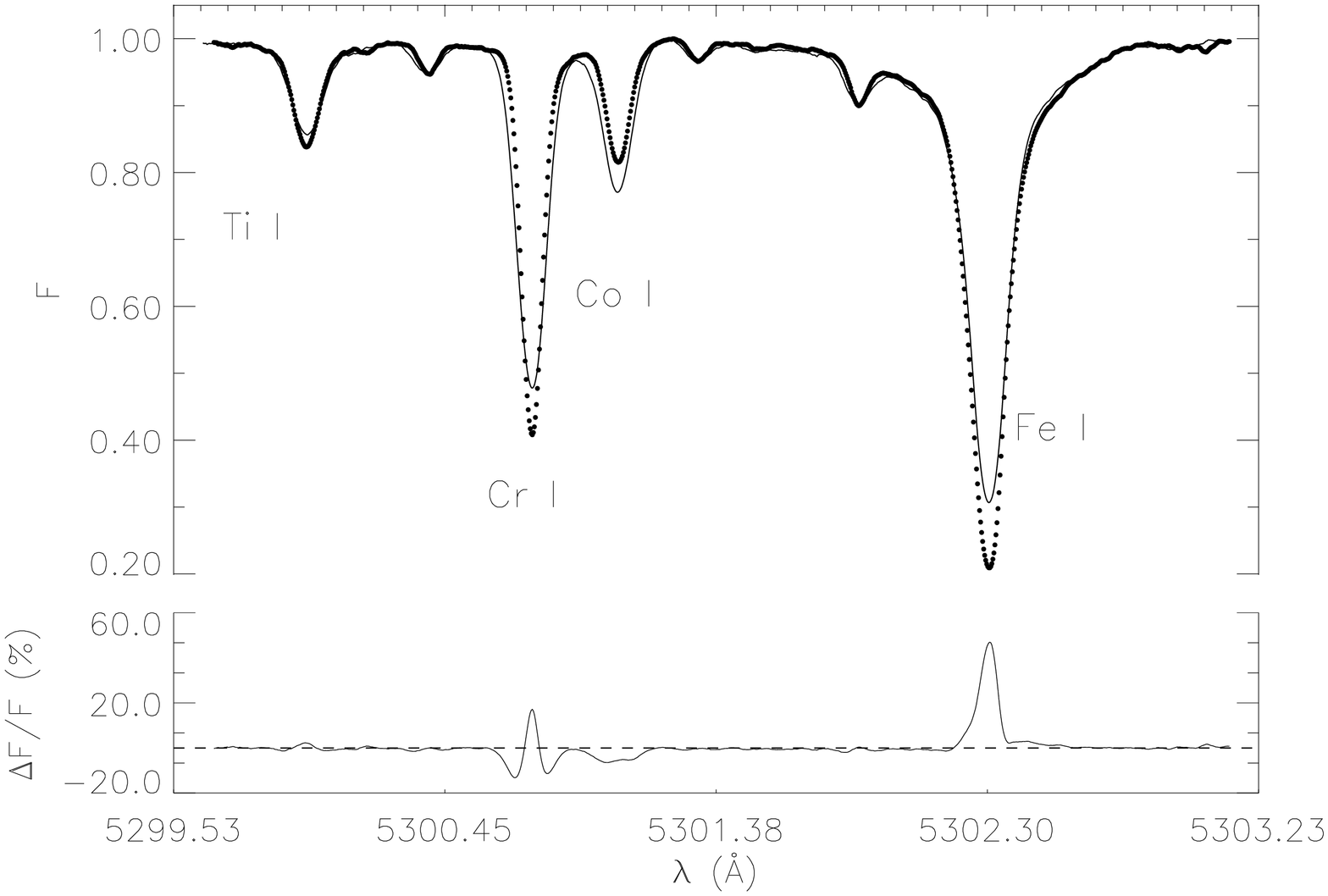}}
\caption{Center (filled circles) and limb (solid) spectra in the region
around Ca I $\lambda6162.2$ (top panel) and around Co I $\lambda 5301.0$ 
(bottom panel).}
\label{f3}
\end{figure}

To avoid a bias due to the
original normalization, the procedure was iterated until 
$ \delta R'/R' < 10^{-3} $. 
Only 3-4 iterations
were required to comply with the convergence condition. Table 1 provides
the results.
Figure  \ref{chi2} shows the contours of the 
reduced $\chi^2$ (normalized to 1 at the minimum) for
the setup centered around the Ca I $\lambda6162$ line. 
The signal in the GCT 
spectra has the potential to exceed a S/N of at least  1000. 
Systematic uncertainties are more difficult to assess.
 Systematic  errors are expected due to shortcomings
of modeling the GCT instrumental profile as a Gaussian, and the scattered
light as independent of wavelength (for a given setup), perhaps with
an additional contribution from real variations of the photospheric
solar spectrum with time. 

 We estimate that our derived values of the resolving
power are good to about a few percent, and that the scattered light is
constrained to within  $\sim$ 1 \%.
Both the relative scattered light, and the polynomial
representing the pseudo-continuum were applied to correct the spectra
at the center of the disk (from which the values were determined)
 and also to the spectra at other positions. Fig. \ref{f2} shows
the FTS spectrum smoothed to the resolution of the 
GCT data,  and the GCT spectrum 
 after correcting the scattered light. We empirically find 
that the observed profiles are accurate to $\sim 1$ \%.
This figure is also supported by an rms difference of $\simeq 0.7$ \% between
two corrected 
spectra of the 6122 \AA\ region at the center of the disk obtained
on the two consecutive observing dates.

Fig. \ref{f3} shows the change from the center to the limb of the
spectral region discussed in the previous figures (upper panel), 
and for the setup centered at $\approx$ 5300 \AA\ (lower panel). 
Many differences are 
noticeable. Some lines only undergo subtle changes,  often 
becoming slightly broader towards the limb, but  
other features' wings turn less pronounced.  
Some lines  become both wider and
stronger towards the limb (e.g. Co  I $\lambda$5301.0). The 
variations in behavior among  different lines 
reflect their different sensitivities to the physical parameters
controlling the line formation and, as earlier recognized in the 60's,
 contain precious information on the atmospheric structure. 
The GCT spectra have been made publicly available through the CDS.
 In the next section we will
explore the possibility of quantifying the thermalizing effect of
inelastic collisions with 
electrons and hydrogen atoms in the populations of neutral 
oxygen and sodium through the center-to-limb behavior 
of spectral lines.

\section{Center-to-limb variation of spectral lines 
and departures from LTE}
\label{clv}

Line profiles sample a range of atmospheric depths that changes depending
on the position on the disk. An analysis of the resulting changes in the
line shapes is, to a first approximation, independent of the $f-$value of the transition, the chemical abundances, and the damping constants.
A zeroth-order analysis can be based on simplified LTE calculations.
Such modeling already reproduces the basic  behavior for many 
spectral lines; their strengthening or weakening from center to limb
due to the changes in the equilibrium populations with atmospheric depth.
A closer look at the observed line profiles will reveal the intricacies 
of the line formation,  such as departures from
LTE,  the presence of inhomogeneities, or the lack of
complete frequency redistribution.  

We explore first the possibility of discerning the thermalizing effect of
collisions with electrons from collisions with hydrogen atoms by means
of an analysis based on one-dimensional model atmospheres and NLTE
line formation. 
When dealing with high S/N and resolving power spectroscopic
observations, an analysis based on static model atmospheres yields 
a very poor match between calculated line profiles
 and observations. This is mainly
the result of neglecting surface inhomogeneities -- granulation-- which
cause the observed lines to appear  asymmetric. 
Admittedly a handicapped approach, such a detour is convenient and
oftentimes didactic.

\begin{figure}[t!]
\centering
{\includegraphics[width=6.3cm,angle=90]{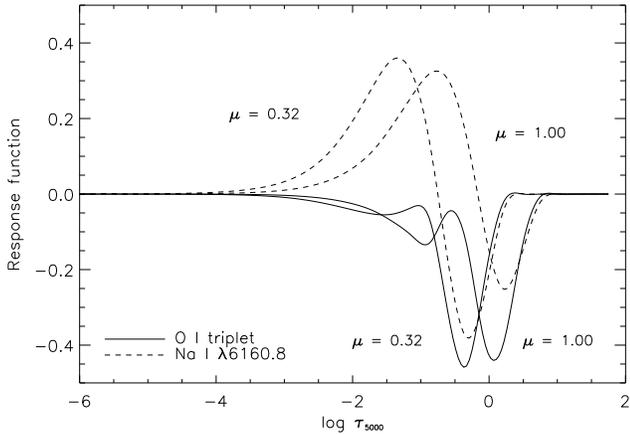}}
\caption{Response functions to the temperature at $\mu=0.32$ and 1.00
for the core of the lines of the O I infrared triplet (solid) and the Na I
$\lambda 6160.8$ (dashed) lines.}
\label{response}
\end{figure}

We examine two interesting cases which are part of
our observations: the oxygen infrared triplet, and a Na I line at 6160.8 \AA.
  Fig.  \ref{response} shows the LTE 
response to temperature perturbations of the core of these lines. 
These response functions (see Ruiz Cobo \& del Toro Iniesta 1992, 1994)
are calculated for the continuum-corrected line profiles, and therefore
include the sensitivity of the continuum, which strengthens the
lines when the temperature increases in the deepest layers. 
In this plot, a positive value represents
an increase in the relative flux, and therefore a weakening of the line.
When changing from the center to the limb, the O I lines map fairly 
well the layers in the range $-0.5 < \log \tau_{\rm 5000} < +0.5$. 
The temperature sensitivity of the core of the Na I line shows
 opposite reactions to changes in the temperature at different
heights but this line is sensitive to higher layers than the triplet.

\begin{figure*}
\centering
{\includegraphics[width=5.5cm,angle=90]{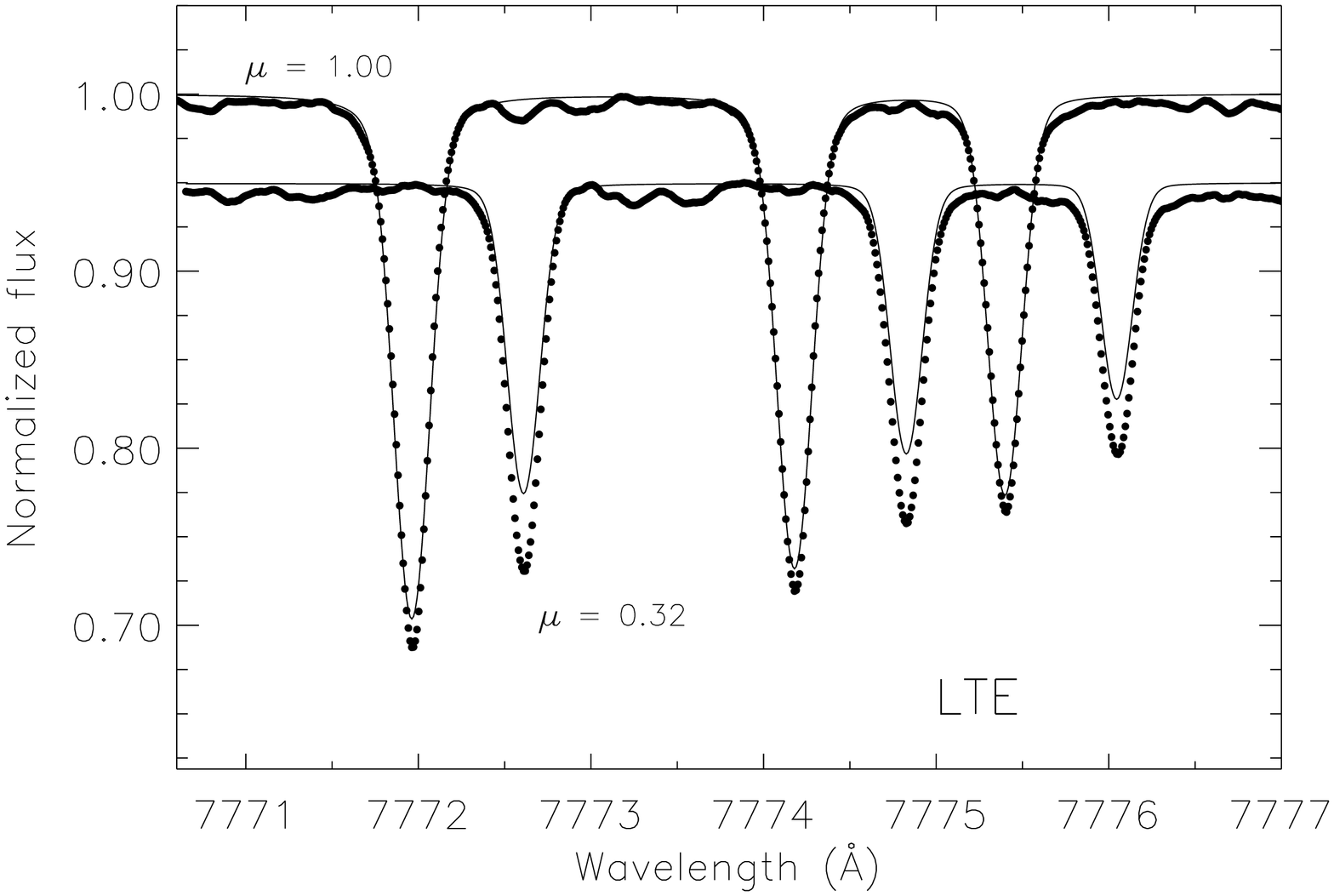}}
{\includegraphics[width=5.5cm,angle=90]{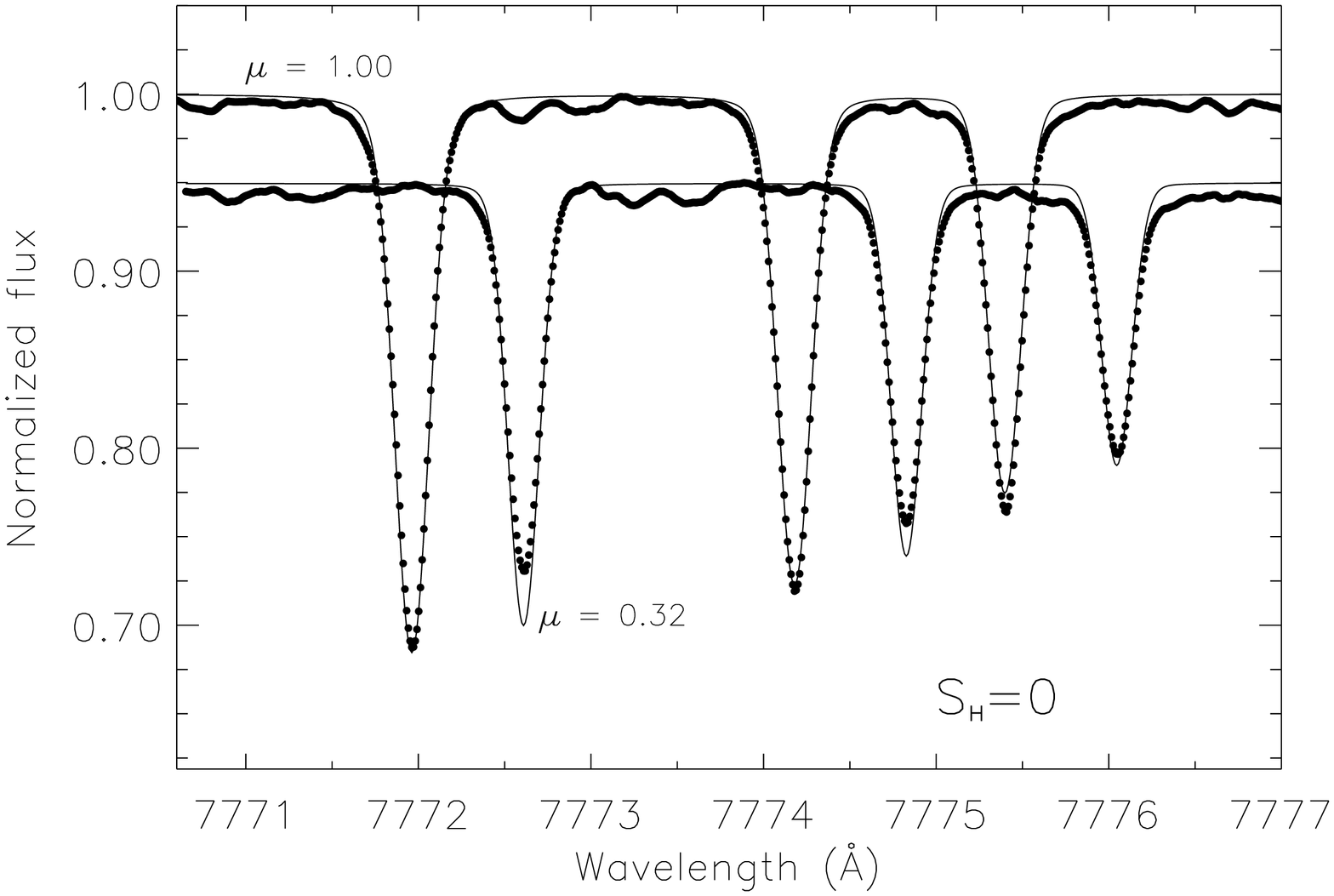}}
{\includegraphics[width=5.5cm,angle=90]{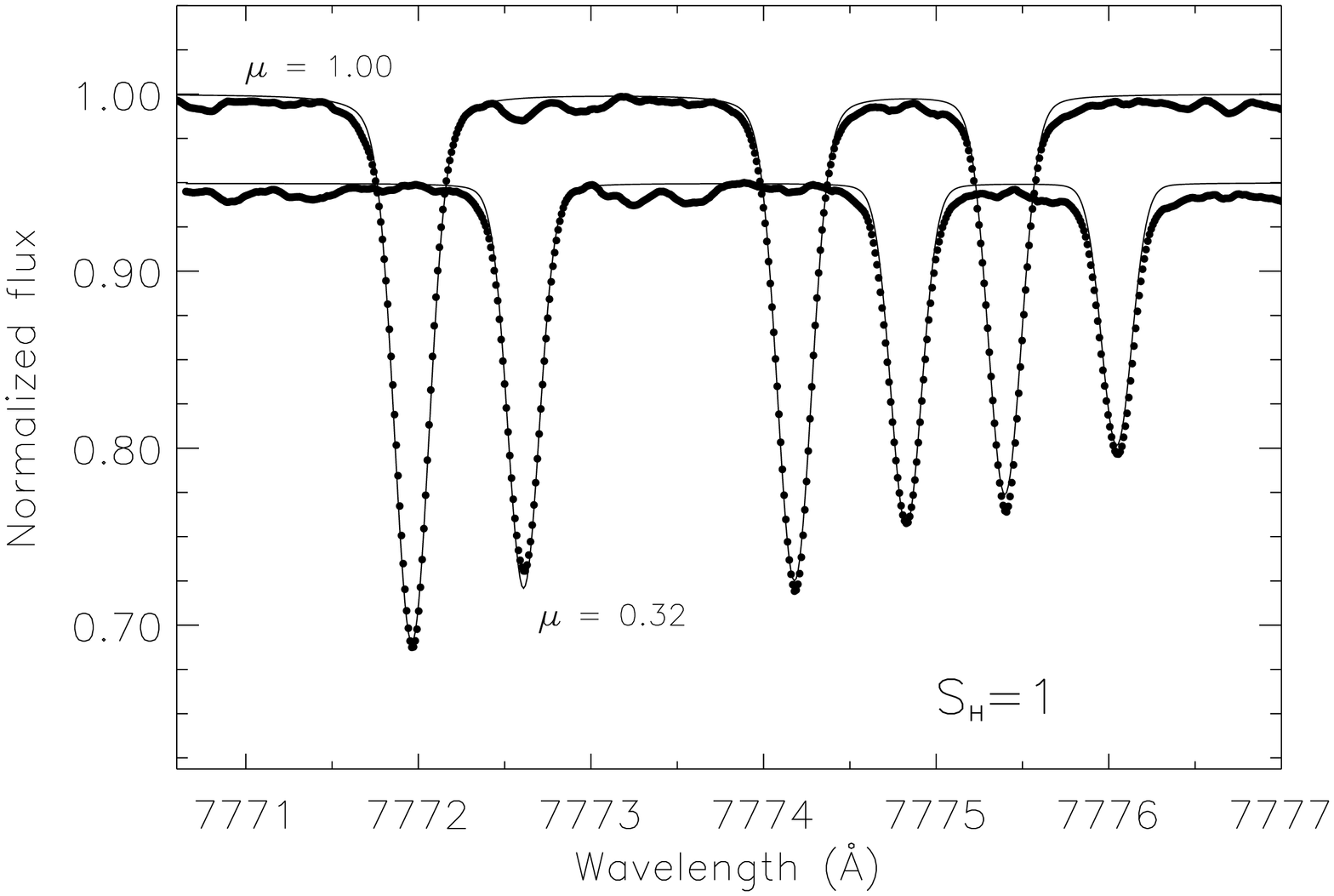}}
{\includegraphics[width=5.5cm,angle=90]{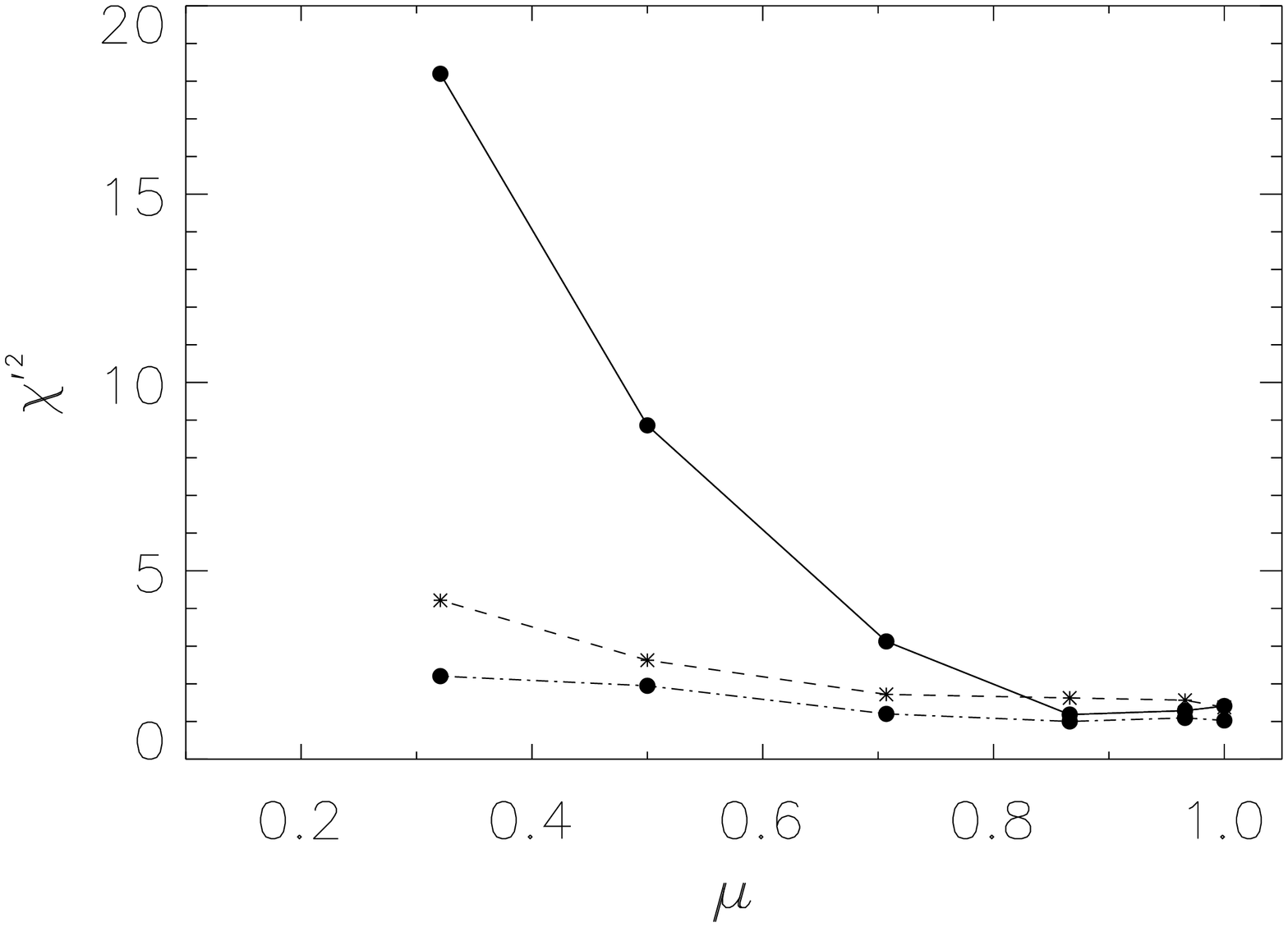}}
\caption{Center (at rest with the continuum normalized  to 1) 
and limb (shifted to the red with the continuum normalized to 1 
and subtracted 0.05) spectra for the O I infrared triplet lines. 
Observations (filled circles) and calculations are compared for three 
different models: LTE, NLTE/$S_{\rm H}=0$, and NLTE/$S_{\rm H}=1$. 
The bottom-right panel shows the variation of the quality of the fit
as a function of the position of the disk: LTE (solid), NLTE/$S_{\rm H}=0$
(dashed) and NLTE/$S_{\rm H}=1$ (dot-dashed). The $\chi'^2$ values have
been normalized to 1 at the minimum.}
\label{7770}
\end{figure*}

\subsection{The oxygen infrared triplet}
\label{oi}

It has been recognized that departures from LTE are significant for these
lines. The main effect is due to an infrared radiation field weaker than Planckian in the line formation region (Eriksson \& Toft 1979). 
Fortunately, the level populations of  O I  are relatively 
insensitive to the UV radiation field in solar type stars, which simplifies
the modeling.
Due to cosmological and galactic implications, oxygen abundances in 
metal-poor stars have been extensively discussed.
When modeling the oxygen triplet lines in NLTE some authors 
opt  to neglect collisions with neutral hydrogen (e.g. Nissen et al. 2002), 
while others prefer to adopt 
a recipe first suggested by Steenbock \& Holweger 
(1984;  see, e.g., Takeda 2003), 
based on Drawin's formula (Drawin 1968), scaled by an empirical factor(s). 

We have evaluated the NLTE populations for O I with a solar model 
 from Kurucz's non-overshooting grid (Kurucz 1993). The model
atom and the calculations follow those in Allende Prieto et al. (2003a, 2003b). 
We have introduced the effect of
collisional excitation and ionization
due to neutral hydrogen with the prescription of
Steenbock \& Holweger, multiplying the rates by a correction 
factor $S_{\rm H}$.
 To make the test independent of the oxygen abundance, 
we have adjusted it for each of the following three cases: 
LTE,  $S_{\rm H} = 0$, and $S_{\rm H}=1$.
A micro-turbulence of 0.9 km s$^{-1}$ was adopted -- a  typical value 
for the disk-center (e.g. Blackwell, Lynas-Gray, \& Smith 1995). 
A Gaussian macro-turbulence of 2.0 km s$^{-1}$ was used, as well a Gaussian instrumental 
profile consistent with the analysis in Section \ref{obs} (see  Table 1),
for a total FWHM broadening of 0.151 \AA. 
Collisional damping
induced by neutral hydrogen was accounted for using the cross-section
at 10,000 K by Barklem, Piskunov \& O'Mara (2000), 
assuming a temperature dependence T$^{2/5}$.

Figure \ref{7770} compares observed and calculated profiles for the
different positions and the three cases considered:
LTE, $S_{\rm H}=0$, and  $S_{\rm H}=1$.
The first three panels illustrate the agreement between observed and 
computed profiles at the extreme positions:
$\mu=1.0$ (center) and $0.32$ (near the limb). The fourth panel quantifies 
the relative agreement through 
$\chi'^2 = \sum (f_{\rm obs} - f_{\rm synth})^2$, which has been
normalized to 1 at the minimum.  To avoid obvious blending 
features, only the fluxes which are more than
3 \% lower than the continuum level are considered in the evaluation 
of $\chi'^2$.
 As collisions drive the level populations toward LTE, the NLTE line profiles   
with $S_{\rm H}=1$ are an intermediate case between LTE and the
$S_{\rm H}=0$  profiles.
 As we anticipated, the match of the observed profiles
with a static one dimensional model is poor.
The comparison suggests that 
the NLTE calculations perform much better than those assuming LTE,
in agreement with previous investigations. We can also conclude that
the effect of hydrogen collisions is quite limited in these lines (see also
Nissen et al. 2002), but the best agreement is found when they are
included, albeit the difference is marginal.

\begin{figure*}
\centering
{\includegraphics[width=5.5cm,angle=90]{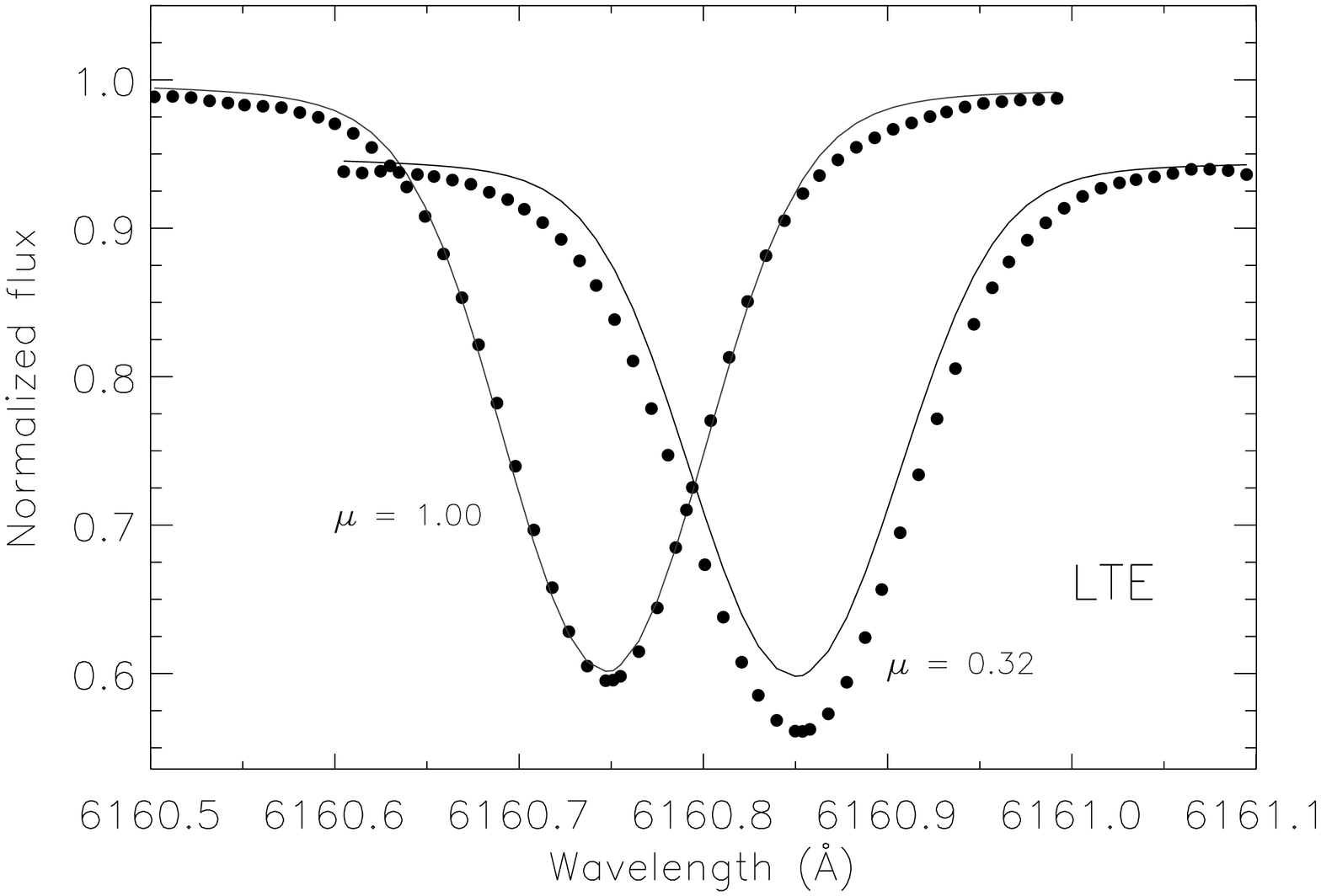}}
{\includegraphics[width=5.5cm,angle=90]{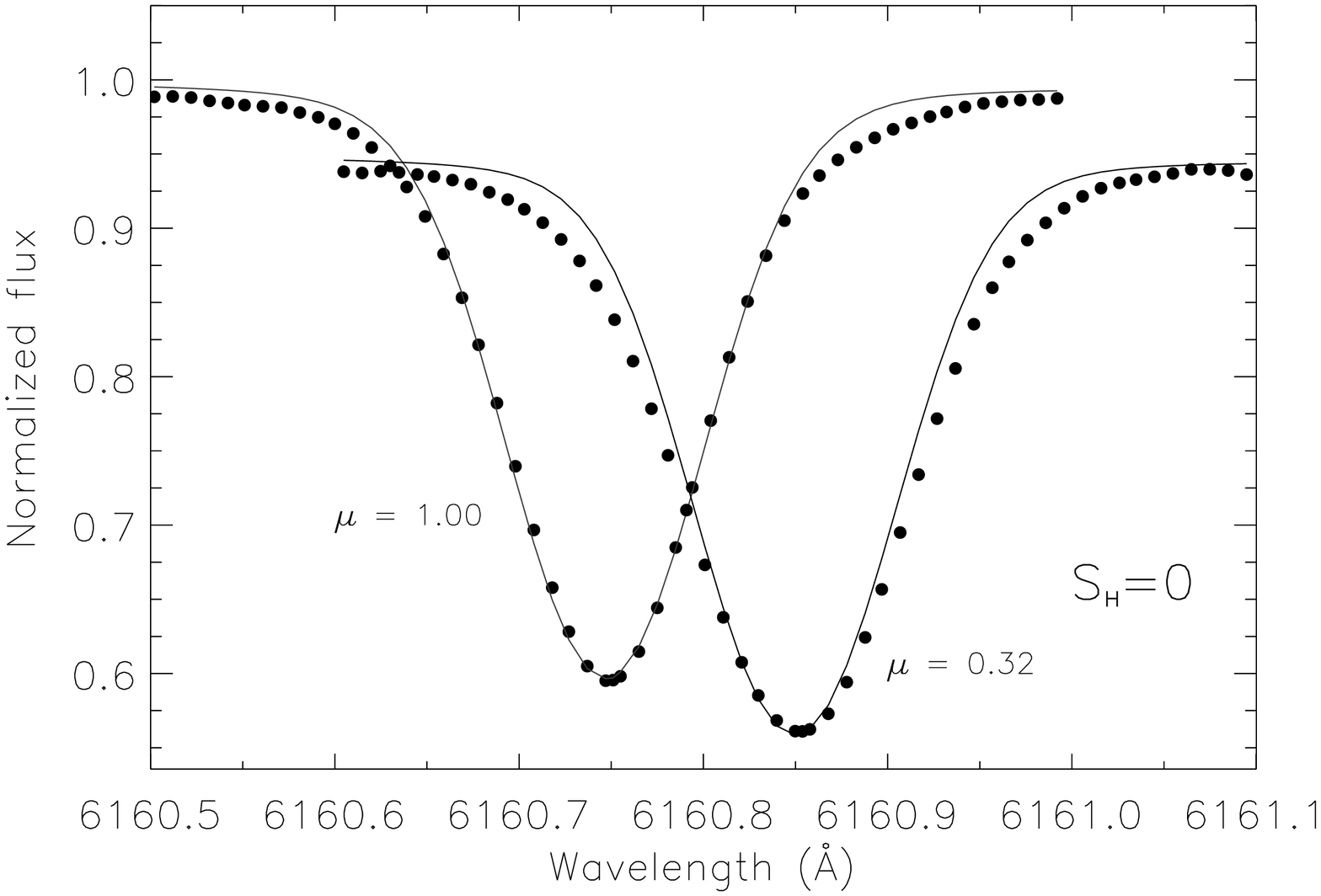}}
{\includegraphics[width=5.5cm,angle=90]{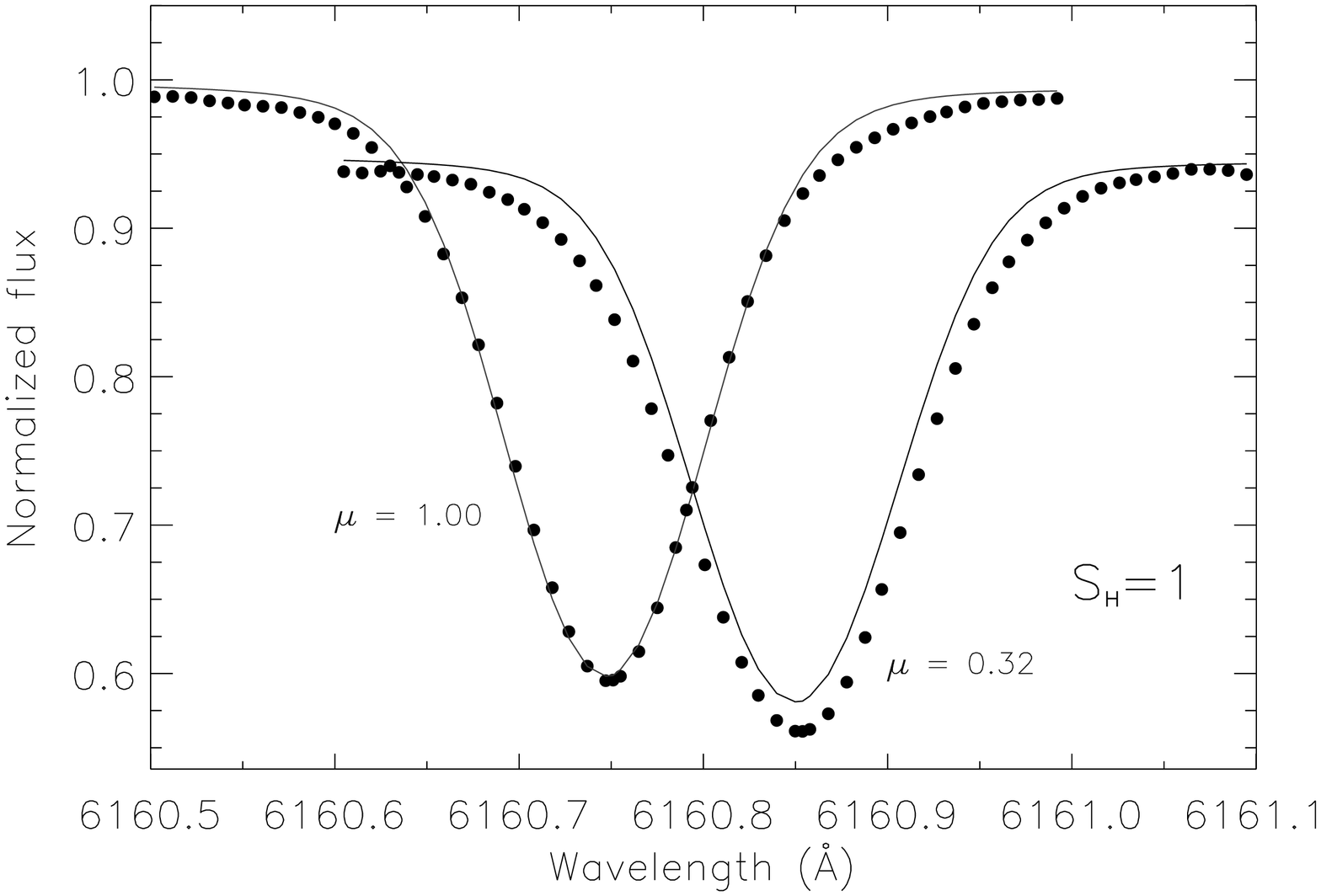}}
{\includegraphics[width=5.5cm,angle=90]{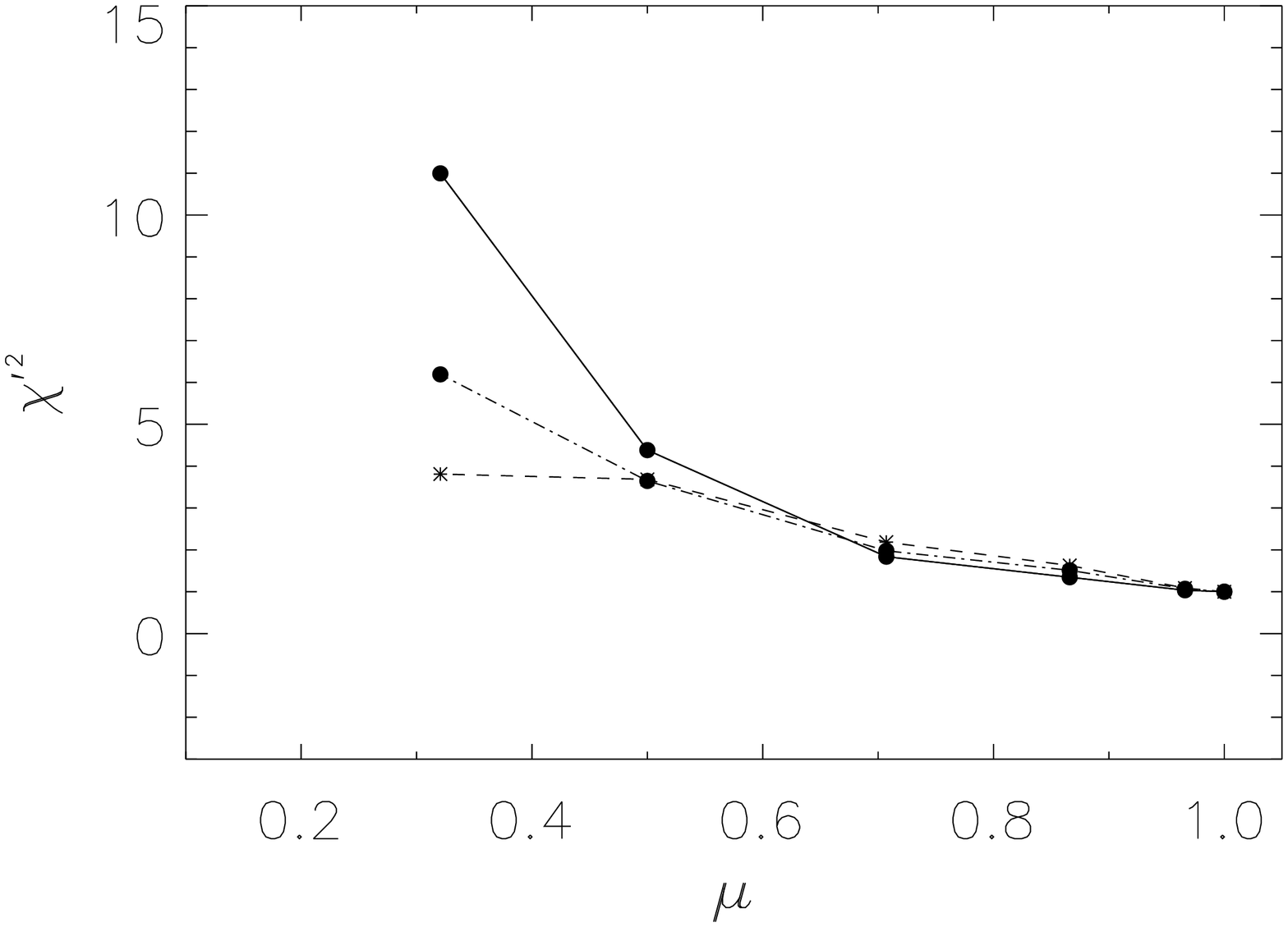}}
\caption{Center (at rest with the continuum normalized  to 1) 
and limb (shifted to the red with the continuum normalized to 1 
and subtracted 0.05) spectra for Na I $\lambda6160.8$. 
Observations (filled circles) and calculations are compared for three 
different models: LTE, NLTE/$S_{\rm H}=0$, and NLTE/$S_{\rm H}=1$.
The bottom-right panel shows the variation of the quality of the fit
as a function of the position of the disk as in Fig. \ref{7770}.}
\label{6161}
\end{figure*}

\subsection{Na I $\lambda 6160.8$}
\label{nai}

Departures from LTE in Na I lines have also been extensively studied 
(e.g. Athay \& Canfield 1969; Baum\"uller, Butler, \& Gehren 1998). 
The level populations are
again  highly independent of the UV radiation field.
We have used the new model
atoms and calculations we previously referred to for the oxygen case. 
A micro-turbulence of 0.9 km s$^{-1}$ was adopted, as in Section \ref{oi},
but a lower value of the macro-turbulence than for the oxygen triplet 
was necessary to fit the profiles at the center of the
disk, 1.3 km s$^{-1}$, and therefore the total FWHM for the Gaussian
broadening was 0.074 \AA\ (note the difference in resolving power,
as summarized in Table 1).  The
effective principal quantum number ($n^{*}$) of the $^2$S upper state 
is larger than 3 and therefore exceeds the range of the 
tables  published by  Anstee and O'Mara (1995).
Barklem et al. (2000) derived damping cross-sections 
for other Na I transitions with similar energies sharing the same
lower state $^2$P$^{\rm o}_{3/2}$. For those transitions 
 the FWHM of the  Lorentzian (collisional) 
component of the line absorption profile 
(per unit perturber number density at 10,000 K)  
$\log \gamma$ was found to be  
in the range $-6.86$ to $-7.23$.  Comparison with the observed 
profiles at the disk center showed that, independently of the 
chosen abundance, the best agreement is found for a value 
close to $\log \gamma = -6.86$, 
which was therefore adopted. Accepting $n^{*} = 3$ for the 
upper $s$ state gives $\log \gamma = -6.96$.

Fig. \ref{6161} shows the profiles at the extreme positions
 ($\mu=0.32$ and 1) and
the comparison with the synthetic spectra for the three cases 
considered. Again, we adjusted the Na abundance for each case in order
to maximize the agreement between calculated and observed profiles 
at the center of the disk.  It is evident that the sensitivity of the level
populations involved in this Na I transition to collisions with H is 
higher than for the states connected by the O I infrared triplet. This
is most likely the result of the line formation and the 
departures from LTE taking place in this case in higher layers. 
The LTE 
response function to the temperature for the core of the O I triplet  
lines peaks at $\log \tau_{\rm 5000} \simeq 0$, while that for
the core of the Na I $\lambda$6160.8 line does so at 
$\log \tau_{\rm 5000} \simeq -1$.
Using the Van Regemorter (1962) and Drawin approximations, 
for a typical transition of a few electron volts 
the ratio of the collisional rate coefficients 
q$_{\rm H}$/q$_{\rm e} \propto (U+2)/U^2 e^{-U}/E_1(U)$, where
$U=E_{ij}/(kT)$, and $E_1$ is the first-order exponential integral. Thus,
q$_{\rm H}$/q$_{\rm e}$ is a fairly flat function of the temperature or
the optical depth in the solar photosphere.
Between $\log \tau_{\rm 5000} \simeq 0$ and $-1$, 
 N$_{\rm H}/$N$_{\rm e}$ increases by an order of magnitude, 
making the effect of collisions with hydrogen more noticeable
in higher layers.
 Similarly to the case of the oxygen triplet, we find the
best agreement with observations when departures from  LTE are considered,
but now the calculations that neglect
collisions with hydrogen grade best, and the differences between
the $S_{\rm H}=0$ and 
1 cases are now more significant. Not even the NLTE $S_{\rm H}=0$ model
can match the wider profile of the sodium line 
that is observed near the solar limb.

\begin{figure*}
\centering
{\includegraphics[width=5.5cm,angle=90]{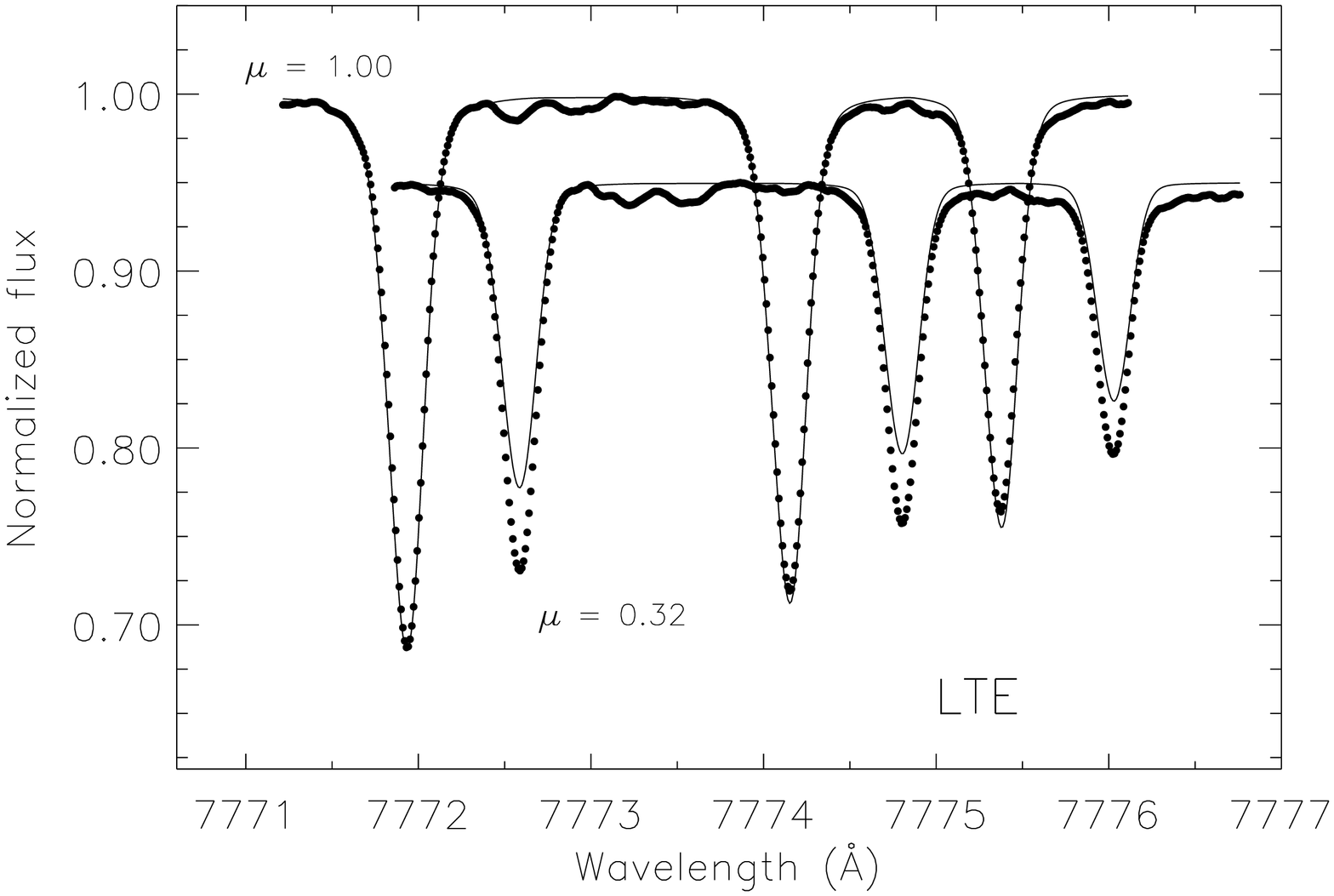}}
{\includegraphics[width=5.5cm,angle=90]{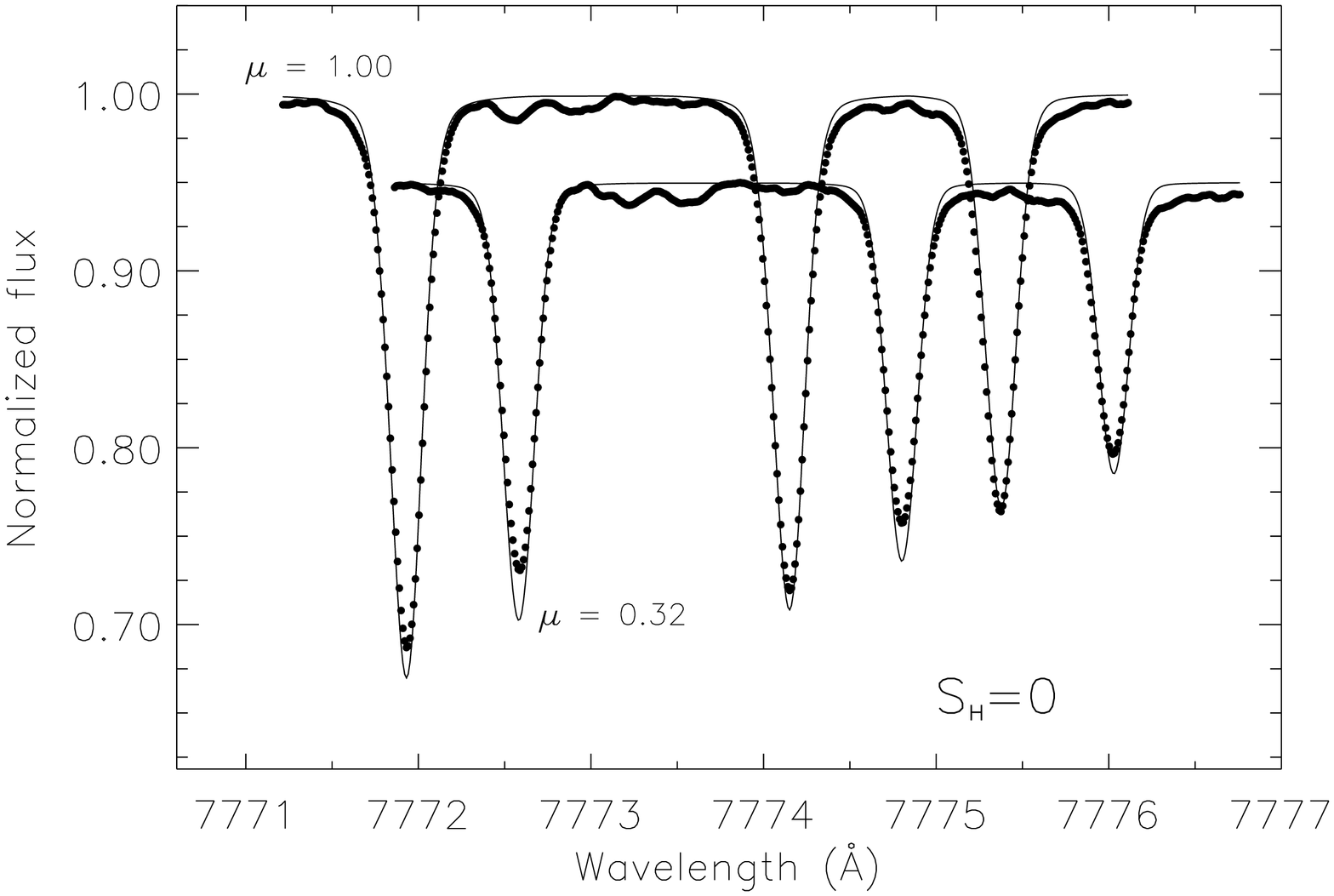}}
{\includegraphics[width=5.5cm,angle=90]{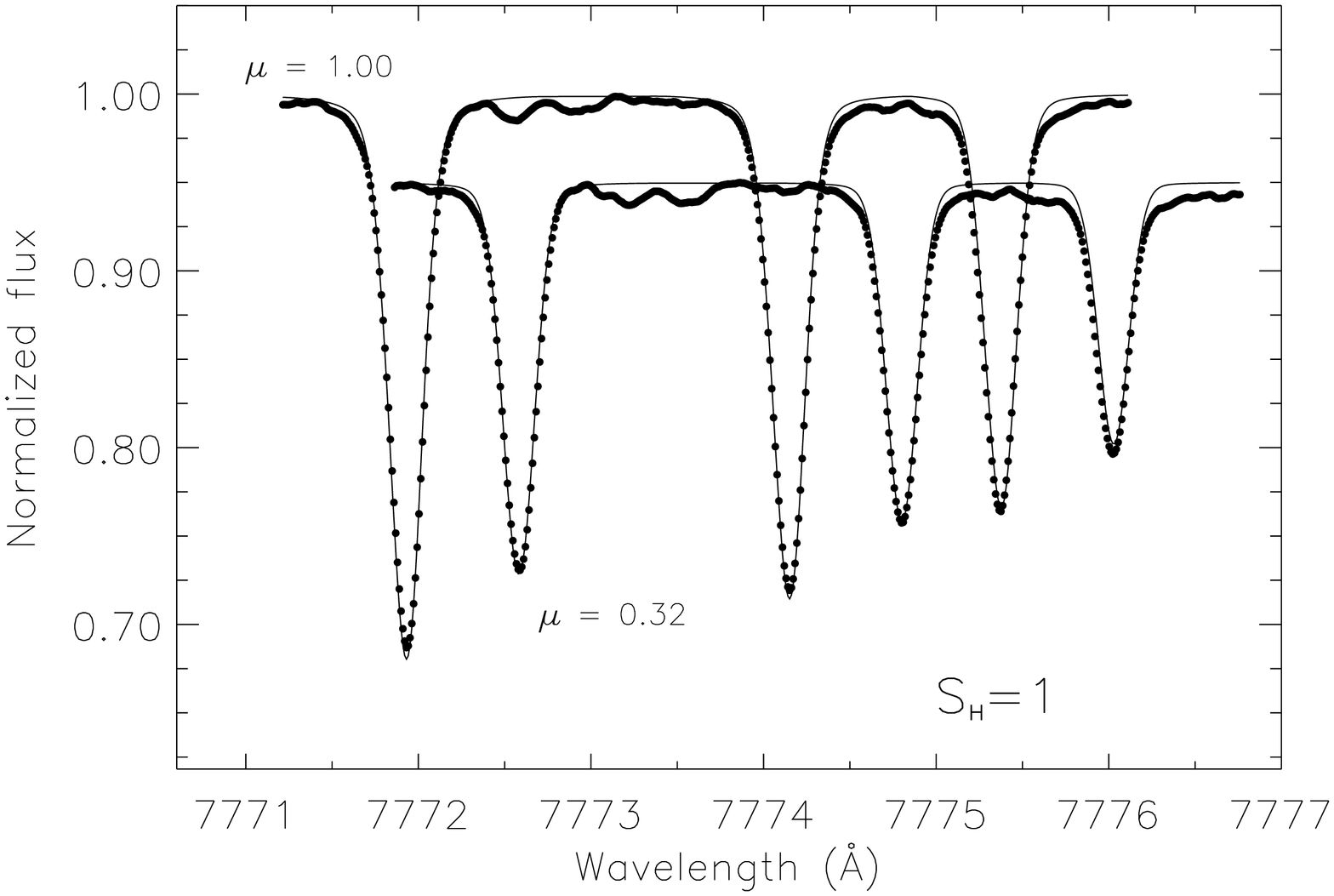}}
{\includegraphics[width=5.5cm,angle=90]{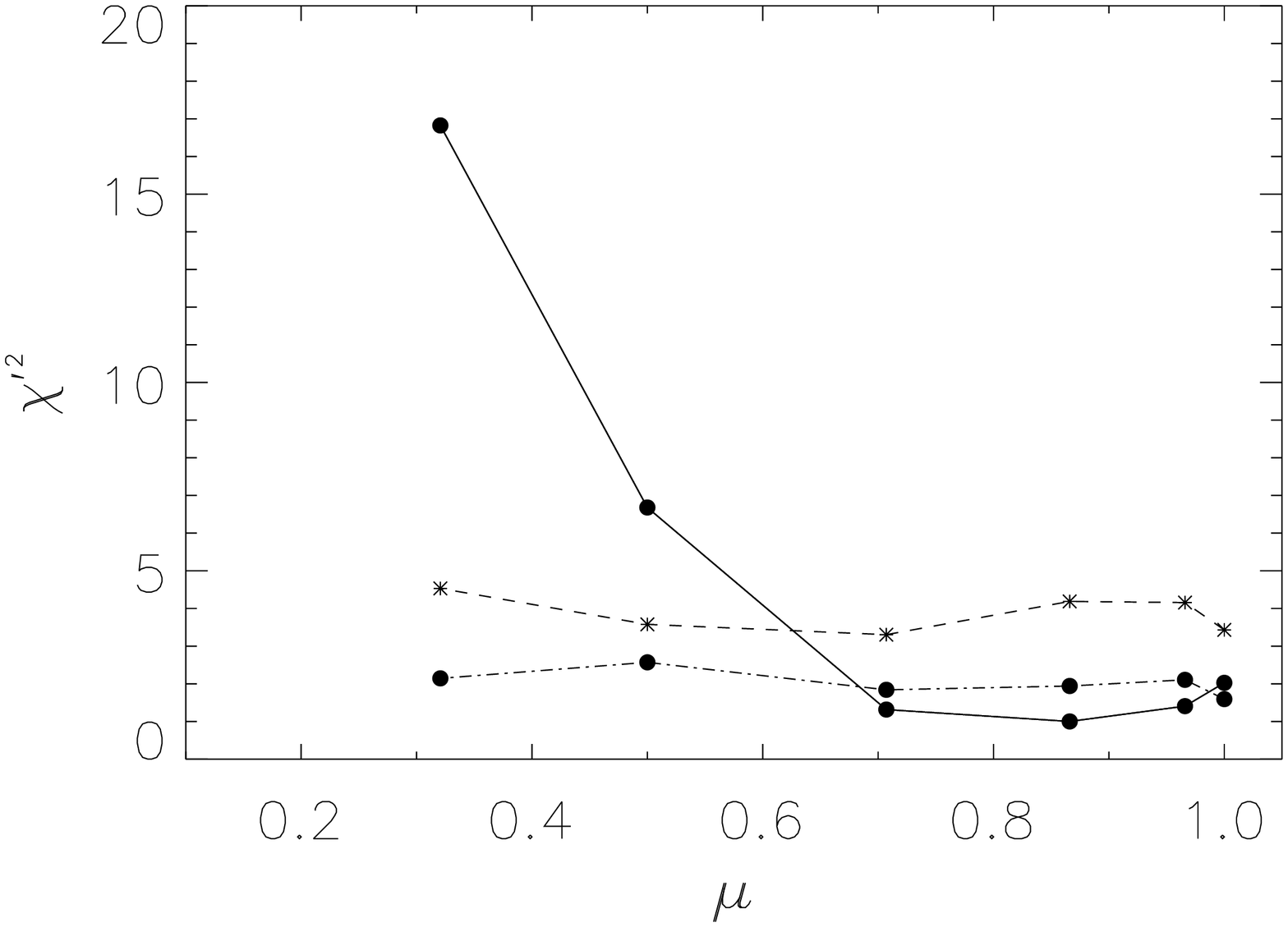}}
\caption{Similar to Fig. \ref{7770} for the 3D time-dependent simulation
of the solar surface. 
}
\label{7770_3D}
\end{figure*}

\section{The O I infrared triplet 
lines in a time-dependent 3D model atmosphere}
\label{3d}

Modern radiation hydrodynamics simulations offer a much more realistic 
representation of the lower solar atmosphere than classical 
static model atmospheres (Stein \& Nordlund 1998; Asplund et al. 2000).
In the context of these, more detailed, model atmospheres, 
surface convection and 
line broadening due to small- and large-scale velocity fields are 
naturally included in the calculations. This makes 
 the use of ad hoc parameters such as macro- and micro-turbulence
unnecessary, and at the same time improves substantially  
the agreement  between observed and calculated spectral line profiles.

The consideration of several more dimensions makes calculations,
however, much more involved, which, with some notable exceptions,
 has largely limited NLTE studies to  1D modeling.
However,   O I  is among the few ions for which the
 line formation in a 3D hydrodynamical model has been  studied 
with a complex multi-level model atom. Asplund et al. (2004)
have recently shown that it is necessary to consider both,  
departures from LTE and a 3D geometry, to find good agreement among
 different spectroscopic indicators of the solar photospheric oxygen 
abundance, finally settling a long-standing problem. Here
we will use  similar calculations to revisit the center-to-limb
variation of these lines, which were analyzed with a 1D model in Section \ref{oi}.

The 3D solar simulation has been described in detail in previous
papers, and  the NLTE line formation calculations are identical to those
reported by Asplund et al. (2004). 
We proceed similarly to the 1D analysis.
Line profiles for the O I triplet  were computed for different
 inclinations from the vertical axis, and
averaged over azimuth, time, and horizontal position. 
The computed NLTE 
profiles are the product of an average LTE profile
derived from 100 snapshots covering 50 minutes of solar time, and
the ratio of the NLTE and LTE line profiles derived from two snapshots.
This procedure reduces the computational burden, while introducing
negligible errors. 
The statistical equilibrium equations were solved for the two considered
snapshots for two different cases: $S_{\rm H} =0$, and $S_{\rm H} =1$.
Three different abundances were used in the calculations
($\log \epsilon$(O) =  8.50, 8.70, and 8.90\footnote{$\epsilon$(O) = 
$10^{12} $N (O)/N(H), where N represents number density.})  
and later the profiles were linearly
interpolated to find the best match at the disk center in each of the
three cases: LTE,  $S_{\rm H} =0$, and $S_{\rm H} =1$. The calculated
profiles were smoothed by convolution with a Gaussian with a FWHM of 
0.088 \AA\ to account for the finite spectral resolution of the observations.

Fig. \ref{7770_3D} is the counterpart of Fig. \ref{7770} for the 3D case.
Admittedly, the computed line profiles are not perfect. In particular,
significant discrepancies are noticeable around the cores of the
lines. 
The reader should keep in mind that the lower-right panels of Figs.
\ref{7770} and \ref{7770_3D}, cannot be directly compared, as the
$\chi'^2$ is normalized to unity at the minimum independently for each case:
1D and 3D. 
The adopted abundances
were chosen to optimize the overall fit of the $\mu=1$ observations, 
and the agreement is considerably
better in 3D than in 1D near the line wings -- even though
 macro-turbulence is included in the 1D case and not in 3D.
The refined 3D analysis   strengthens what the  1D study 
suggested. The calculated
profiles are now close enough to the observations to justify an attempt
to quantify statistically the agreement. 
Making use of the accuracy estimate empirically derived in Section \ref{obs}, we adopt $\sigma \simeq 0.01$, 
and evaluate 
$\chi^2 = 1/\sigma^2 
\sum_{\rm i=1}^{\rm n} (f_{\rm obs} - f_{\rm synth})^2$  for 
$\mu= 0.32$ in the three cases. Using only 
the fluxes below 3\% of the continuum level to avoid the interference
of obvious blending features between the lines, 
 n$=135$, and we find $\chi^2$= 725, 195, and 92
and therefore significance levels of 0 ($< 10^{-81}$),  
10$^{-3}$, and $1$, for LTE, $S_{\rm H}=0$, and $S_{\rm H}=1$, respectively. 

In the 1D analyses we have avoided comparing absolute abundances, given
that the effect of atmospheric inhomogeneities was neglected. 
Now that the triplet lines are modeled in more detail, it is tempting
to examine the absolute abundances as
an additional piece of evidence. The abundances found in each case to 
match the profiles at the
center of the disk are $\log \epsilon$(O) = 8.88, 8.66 and 8.72 for  
LTE, $S_{\rm H}=0$, and $S_{\rm H}=1$, respectively. The 
$S_{\rm H}=0$ case provides the best agreement 
with the average value 
derived by Asplund et al. (2004)\footnote{
Our derived oxygen abundance for the $S_{\rm H}=0$ case  (8.66 dex) 
is consistent with the abundances derived by Asplund et al. (2004) by
fitting the profiles of the same lines in the solar flux atlas 
of Kurucz et al. (1984): 8.64--8.66 dex
}: $8.66 \pm 0.05$ dex from forbidden and
permitted lines of O I and from infrared OH lines. 
Thus, absolute abundance and $\chi^2$ statistics 
lead to apparently opposite conclusions in the case of the oxygen triplet.
We should note that the average value derived by Asplund et al. (2004)
includes permitted lines analyzed in a similar manner as in this 
paper with NLTE line formation and $S_{\rm H}=0$. A higher oxygen 
abundance from allowed lines is expected for $S_{\rm H}=1$, and in that
 case, forbidden and permitted lines
would suggest a higher abundance (8.67--8.70) than infrared 
OH lines (8.61--8.65) dex.

We should stress that the necessary atomic data for the 
NLTE calculations were 
independently compiled for the 1D and 3D cases, yet the qualitative 
agreement is excellent.  Quantitatively, the abundance corrections 
necessary to make the LTE and NLTE equivalent widths agree 
are also remarkably similar in 1D and 3D.
This was also found by Asplund et al. (2004)  comparing the 3D 
calculations with 1D abundances based on 
MARCS model atmospheres. The impact of
departures from LTE and that of surface inhomogeneities 
on the abundances cannot be generally decoupled,
but the situation for the OI triplet, where this is actually
a good approximation, makes the conclusions from our 1D analysis to
be similar as those from a 3D study.

\section{Summary and conclusions}

The type of observations presented here, as exemplified in Sections \ref{clv} 
and \ref{3d}, 
provide a stiff test of theoretical calculations
 of atmospheric structure and line formation in a solar-type photosphere.
The potential of center-to-limb observations  to guide
 modeling was recognized early, but 
consistent $\mu \neq 1$ observations covering extensive spectral regions are
still missing. 
We have explored how these observations can constrain
the  effect of collisions with hydrogen atoms on the rate
equations for two particular cases.

Our results for Na I $\lambda$6160.8  
 fall in line with 
the conclusion drawn from theoretical calculations 
by Barklem, Belyaev, \& Asplund (2003)  for excitation of another alkali,
 lithium, by inelastic collisions with H atoms (see also the discussion by
Lambert 1993). 
Limited by the accuracy of the observations, and approximations
involved in computing line profiles,
we have only tested here the extreme cases of $S_{\rm H}=0$ (no collisions
with H), or $S_{\rm H}=1$ (Drawin-like formula).
Comparison  with detailed NLTE calculations based on 
multidimensional time-dependent model atmospheres and improved solar 
observations free from scattered light may be able to answer whether
or not the use of a simplified formula scaled by a constant factor 
($<<1 $ for Na I or Li I) is a useful approach. 
Our results and a quick inspection of the literature 
lead to the naive conclusion that such a factor would need to 
vary for different species.
For example, 
the role of hydrogen collisions compared to those with electrons 
 increases for atmospheres more metal-poor (or cooler) than solar, and
Korn, Shi, \& Gehren (2003) have recently found that 
$S_{\rm H}=3$ is needed to satisfy the iron ionization balance in
several metal-poor stars.

In the case of the oxygen infrared triplet, 
the solar observations are best reproduced considering inelastic
hydrogen collisions in the rate equations.
However, the oxygen abundance
obtained when hydrogen collisions are neglected (8.66 dex) 
is in better agreement
with the abundance inferred from OH infrared lines (8.61--8.65 dex;
Asplund et al. 2004) than
the abundance derived from the NLTE $S_{\rm H}=1$ analysis (8.72 dex).
The higher $S_{\rm H}=1$ abundance is nevertheless still consistent with the 
values from forbidden lines (8.67--8.69 dex).
The weighted average of the abundances from permitted lines derived
by Asplund et al. (2004) changes from $8.64 \pm 0.02$ dex 
to $8.70 \pm 0.04$ dex when inelastic collisions with hydrogen are
considered with the Drawin-like formula. Therefore, this
apparent inconsistency could signal that the abundance derived
from infrared OH lines is more uncertain than the values
obtained from a detailed analysis of forbidden and permitted O I lines.

We should bear in mind that 
the effect of collisions with hydrogen on the statistical equilibrium is
not the only important uncertainty for NLTE calculations in late-type
stellar atmospheres. Electron collisions, for example, 
 are only approximately included
in our models, mainly due to the lack of reliable data for particular states.
Our results should be confirmed by  taking 
into account refined data as they become available 
(see, e.g., Zatsarinny \& Tayal 2003). 
In addition, the electron density in high atmospheric layers is highly
uncertain even in the 3D hydrodynamical simulations, due to the assumption
of LTE. Our NLTE calculations are always `restricted', in the sense that
the temperature and electron density 
are adopted from the LTE structure calculations.

The examples discussed in this paper show how the different ingredients
involved in modeling spectral line formation can begin to 
be unraveled  when high-quality observations of 
spatially resolved solar line profiles
 are compared with detailed calculations.  High quality 
solar observations with low spatial resolution are still missing at
most wavelengths and can play a crucial role guiding theory 
in the quest for accuracy in the interpretation of stellar spectra.

\begin{acknowledgements}

It is a pleasure to thank Ram\'on Garc\'{\i}a L\'opez for encouraging 
  us to perform these observations, and 
Valent\'{\i}n Mart\'{\i}nez Pillet for allocating the telescope time.
We are grateful to Dan Kiselman for pointing out an important mistake
in an earlier version of the manuscript, and to the referee, Han Uitenbroek,
and the scientific editor, Wolfgang Schmidt, for useful suggestions.
The Gregory Coud\'e Telescope was operated by the
 Universit\"ats-Sternwarte G\"ottingen at the Spanish Observatorio del
Teide of the Instituto de Astrof\'{\i}sica de Canarias.
We  made use of NASA's ADS, and gratefully acknowledge support from 
NSF (grant AST-0086321) and NASA (LTSA 02-0017-0093).

\end{acknowledgements}

\end{document}